\RequirePackage{fix-cm}
\documentclass[english,prl,reprint,twocolumn,superscriptaddress]{revtex4-1}
\usepackage[T1]{fontenc}
\usepackage[latin9]{inputenc}
\usepackage{xcolor}
\usepackage{babel}
\usepackage{mathrsfs}
\usepackage{amsbsy}
\usepackage{amstext}
\usepackage{amssymb}
\usepackage{graphicx}
\usepackage[unicode=true,pdfusetitle,
 bookmarks=true,bookmarksnumbered=false,bookmarksopen=false,
 breaklinks=true,pdfborder={0 0 0},pdfborderstyle={},backref=false,colorlinks=true]
 {hyperref}
\hypersetup{
 colorlinks,citecolor=blue,linkcolor=blue,urlcolor=blue,bookmarks=false,hypertexnames=true}
\begin{document}
\global\long\def\vect#1{\overrightarrow{\mathbf{#1}}}%

\global\long\def\abs#1{\left|#1\right|}%

\global\long\def\av#1{\left\langle #1\right\rangle }%

\global\long\def\ket#1{\left|#1\right\rangle }%

\global\long\def\bra#1{\left\langle #1\right|}%

\global\long\def\tensorproduct{\otimes}%

\global\long\def\braket#1#2{\left\langle #1\mid#2\right\rangle }%

\global\long\def\omv{\overrightarrow{\Omega}}%

\global\long\def\inf{\infty}%

\title{Anomalous Transport Signatures in Weyl Semimetals with Point Defects}
\date{\today}
\author{J. P. Santos Pires}
\email{up201201453@fc.up.pt}

\address{Departamento de Física e Astronomia, Faculdade de Ciências da Universidade
do Porto, Rua do Campo Alegre, s/n, 4169-007 Porto, Portugal}
\address{Centro de Física das Universidades do Minho e do Porto (CF-UM-UP)
and Laboratório de Física para Materiais e Tecnologias Emergentes
LaPMET, University of Porto, 4169-007 Porto, Portugal}
\author{S. M. João}
\address{Departamento de Física e Astronomia, Faculdade de Ciências da Universidade
do Porto, Rua do Campo Alegre, s/n, 4169-007 Porto, Portugal}
\address{Centro de Física das Universidades do Minho e do Porto (CF-UM-UP)
and Laboratório de Física para Materiais e Tecnologias Emergentes
LaPMET, University of Porto, 4169-007 Porto, Portugal}
\author{Aires Ferreira}
\address{Department of Physics and York Centre for Quantum Technologies, University
of York, YO10 5DD, York, United Kingdom}
\author{B. Amorim}
\address{Centro de Física das Universidades do Minho e do Porto (CF-UM-UP)
and Laboratório de Física para Materiais e Tecnologias Emergentes
LaPMET, Universidade do Minho, Campus de Gualtar 4710-057 Braga, Portugal}
\author{J. M. Viana Parente Lopes}
\email{jlopes@fc.up.pt}

\address{Departamento de Física e Astronomia, Faculdade de Ciências da Universidade
do Porto, Rua do Campo Alegre, s/n, 4169-007 Porto, Portugal}
\address{Centro de Física das Universidades do Minho e do Porto (CF-UM-UP)
and Laboratório de Física para Materiais e Tecnologias Emergentes
LaPMET, University of Porto, 4169-007 Porto, Portugal}
\begin{abstract}
We present the first theoretical study of transport properties of
Weyl semimetals with point defects. Focusing on a class of time-reversal
symmetric Weyl lattice models, we show that dilute lattice vacancies
induce a finite density of quasi-localized states at and near the
nodal energy, causing strong modifications to the low-energy spectrum.
This generates novel transport effects, namely\textit{\textcolor{black}{{}
}}\textit{\textcolor{black}{\emph{(i)}}}\textit{ }an oscillatory behavior
of the dc-conductivity with the charge carrier density in the absence
of magnetic fields, and\textit{\textcolor{black}{{} }}\textit{\textcolor{black}{\emph{(ii)}}}
a plateau-shaped dissipative optical response for photon frequencies
below the inter-ban\textcolor{black}{d threshold, $E_{F}\!\lesssim\!\hbar\omega\!\lesssim\!2E_{F}$.
Our resul}ts provide a path to engineer unconventional quantum transport
effects in Weyl semimetals by means of common point defects.
\end{abstract}
\maketitle
Weyl semimetals (WSMs) are a new class of topological materials, whose
low-energy excitations behave as chiral Weyl fermions in (3+1) dimensions\,\citep{Armitage2018}.
Their nontrivial topology, resulting from the separation of Weyl nodes
with opposite chirality in momentum space, finds one of its prime
manifestations in unusual surface states called Fermi arcs\,\citep{Wan2011,Witten2015,Haldane2014,Hashimoto2017,Qian_21}.
Because WSMs provide a route towards the realization of the chiral
anomaly, they have become a bedrock for the correspondence between
quantum field theory and condensed matter physics\,\citep{Zyuzin_12,Hosur2012,Son2013,Zhang2016,Barnes_16,Ong2021}.

WSMs likewise provide a fertile ground for exploring the interplay
of band topology and disorder effects, both from the perspective of
quantum phase transitions at zero temperature \citep{Kobayashi2014,Pixley2015,Bera2016,Roy18}
and key transport signatures of the chiral anomaly, such as the chiral
magnetic effect \citep{Burkov_14,Lu_15,Burkov_17}. Moreover, theoretical
studies indicate that the inclusion of smooth disordered potentials
preserves the topological protection enjoyed by clean WSMs, rendering
well-decoupled Weyl nodes\,\citep{Ominato_14,Altland_15,Altland_16}
connected by robust surface Fermi-arcs\,\citep{Chen15,Liu16,Brillaux21}.

Meanwhile, the impact of realistic disorder landscapes beyond the
standard mean-field picture has proved to be a subtle and intriguing
problem \citep{Pixley2016,Buchhold2018}. The inevitable existence
of so-called \emph{rare regions} within a disorder landscape, in which
the scattering potential attains a constant value, was found to yield
a small, \emph{but nonzero}, density of states (DoS) at the nodal
energy \citep{Nandkishore2014,Pixley16a,Wilson2020,Pixley2021,Pires2021}.
\textcolor{black}{Rare fluctuations of the disorder potential induce
power-law localized resonances within the continuum \citep{Nandkishore2014}
(not unlike the Lifshitz-tail phenomenon \citep{Yaida16,Gurarie2017}),
which are manifest in weakly disordered samples with arbitrarily long
elastic scattering lifetimes and statistically relevant in the thermodynamic
limit \citep{Pires2021}}. Notwithstanding, rare-region events produce
a minute effect on measurable quantities (the nodal DoS lifting is
exponentially small in the inverse disorder strength\,\citep{Nandkishore2014,Pixley16a}),
which makes them very challenging to detect. With this in mind, this
Letter discusses alternative mechanisms that can yield \emph{model-free,
distinctive signatures of resonant zero-energy states amenable to
experimental verification}. We posit that vacancies are ideal candidates
for this purpose because they are common point defects that act as
strong scatterers of charge carriers whilst preserving the non-spatial
symmetries of the underlying lattice model \citep{Pereira08,Ostrovski2014,Hafner2014,Ferreira2015}.
Moreover, vacancies can be intentionally created by light-ion irradiation\,\citep{Zhang2019,Fu2020}
and have been shown to generate \textsl{scale-free bound states} at
the nodal energy of WSMs\,\citep{Pires2022}, thus providing a practical
route to explore transport signatures of emergent zero-energy modes
under controlled conditions. \textcolor{black}{Point defects are expected
to be chiefly important in WSMs of the TaAs family, including NbAs,
TaP and NbP, which typically crystalize as cubic lattices. These can
be synthesized by standard chemical vapor transport techniques\,\citep{Ghimire2015}
and are experimentally known to host a significant concentration of
point defects, even in the highest quality crystals\,\citep{Besara2016,Liu2016}.}

Here, we report new real-space simulations of charge carrier transport
and optical response in a time-reversal symmetric WSM hosting a finite
concentration of point defects. Our study reveals several novel transport
effects. First, the electrical conductivity is shown to display an
\emph{oscillatory behavior} on top of the standard monotonic variation
with the charge carrier density. This is in stark contrast to the
monotonic dependence in analogous 2D Dirac models \citep{Ferreira2015},
and cannot be easily replicated by other disorder species. Rather,
it derives from strong inter-vacancy interference effects that efficiently
modulate the response of the Fermi surface to external fields in slightly
doped WSMs. Second, and equally important, the optical response of
WSMs with dilute vacancy defects exhibits uniform absorption over
a wide frequency rang\textcolor{black}{e, $E_{F}\lesssim\hbar\omega\lesssim2E_{F}$,
}below the onset of inter-band transitions. This effect is a unique
manifestation of emergent nodal bound states amenable to experimental
verification. Finally, additional numerical results on the electronic
structure are also shown with the aim of deciphering vacancy-induced
zero energy modes and characterizing their sensitivity to applied
magnetic fields.

\emph{Electronic structure}.---The WSM is described by a two-orbital
tight-binding model defined on a simple cubic lattice ($\mathcal{L}$),

\vspace{-0.5cm}
\begin{equation}
\mathcal{H}_{0}\!=\!\frac{\hbar v}{2a}\sum_{\mathbf{R}\in\mathcal{L}}\sum_{{\scriptscriptstyle i=x\!,\!y\!,\!z}}\!\!\left[\Psi_{{\scriptscriptstyle \mathbf{R}}}^{\dagger}\!\cdot\!\sigma^{i}\!\!\cdot\!\Psi_{{\scriptscriptstyle \mathbf{R}+a\mathbf{x}_{i}}}\!-\!\text{h.c.}\right],\label{eq:HamiltonianLattice}
\end{equation}

\vspace{-0.2cm}

\noindent where $a$ is the lattice parameter, $v$ is the Fermi velocity,
$\mathbf{x}_{i}\!=\!(\hat{x},\hat{y},\hat{z})$ are the cartesian
unit vectors, $\boldsymbol{\sigma}$ is a vector of $2\!\times\!2$
Pauli matrices, and $\Psi_{{\scriptscriptstyle \mathbf{R}}}^{\dagger}\!=\![c_{\mathbf{R},1}^{\dagger},c_{\mathbf{R},2}^{\dagger}]$
is a local two-orbital fermionic creation operator\,\citep{Pixley2021}.
A \emph{full-vacancy} acts as a local perturbation that removes all
hoppings between orbitals at the defect site and its neighbors. In
a companion paper\,\citep{Pires2022}, we have shown that one such
point defect produces a bound state at the nodal energy, whose real-space
wavefunction decays asymptotically with an inverse-square law. In
what follows, we employ large-scale Chebyshev expansions of lattice
Green's functions\,\citep{Weise2006,Ferreira2015}, as implemented
in the KITE code\,\citep{Joao2020}, to determine the impact of vacancy-induced
nodal states on several quantities of interest.

\noindent \ \ \ We start by calculating the change to the thermodynamic
DoS induced by a finite concentration of randomly placed vacancies,
$n_{v}$. We recall that the clean DoS vanishes quadratically at the
nodal energy and, therefore, a pristine WSM realizes an incompressible
electronic phase in the absence of external fields. Unsurprisingly,
symmetry breaking due to disorder will change this picture by transferring
spectral weight across the energy spectrum, with previous studies
of white-noise scalar potentials and extended impurities showing that
the significant changes in the DoS occur away from the nodal energy,
owing to the topological protection enjoyed by Weyl fermions\,\citep{Nandkishore2014,Pixley16a,Wilson2020,Pixley2021,Pires2021}.
The situation with vacancies is strikingly different. If random vacancies
within a sample were taken in isolation, the single-vacancy result
of Ref.\,\citep{Pires2022} would imply that a spectral weight proportional
to $n_{v}$ is drawn out of the continuum and placed exactly at the
nodal energy. Generally, such situation cannot be maintained for sufficiently
large defect concentrations and, in fact, coherent multiple-scattering
may become important in the quantum regime, even at low concentrations\,\citep{Ferreira2015,Joao2021}.
\begin{figure}[t]
\begin{centering}
\hspace{-0.2cm}\includegraphics[scale=0.24]{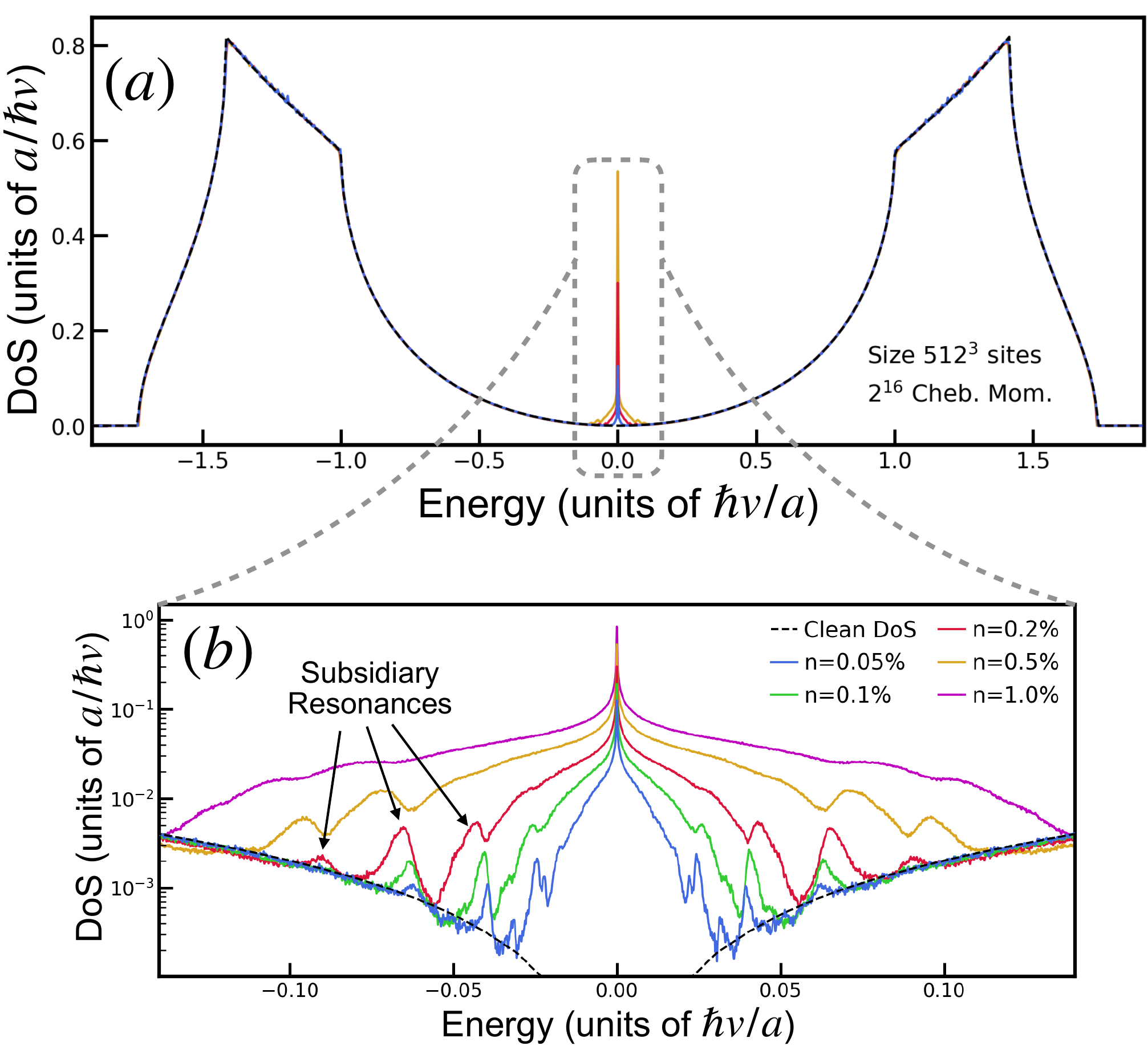}
\par\end{centering}
\vspace{-0.3cm}

\caption{\label{fig-01}Average DoS of a large WSM lattice, with linear size
$L\!=\!512a$, for several vacancy concentrations. An overview of
the entire spectrum is presented in \textit{(a)} and a close-up around
the central peak is shown in \textit{(b)}. The comb of subsidiary
resonances is marked by arrows in \textit{(b)}. The spectral resolution
of the calculations was set to $10^{-3}\,\hbar v/a$.}

\vspace{-0.7cm}
\end{figure}

\noindent \ \ \ To reliably capture quantum coherence effects,
we consider large cubic systems with linear size $L$. The DoS is
obtained by simulating a system with $L\!=\!512a$, using an exact
Chebyshev expansion of the resolvent operator convoluted with a Jackson
damping kernel\,\citep{Weise2006}. Finite size effects are eliminated
by jointly averaging over defect configurations and random twisted
boundary conditions, which yields virtually exact simulations within
resolution. Our results, summarized in Fig.\,\ref{fig-01}, disclose
a strong enhancement in and around the node for any nonzero vacancy
concentration. As anticipated, this sharply contrasts with the case
of a random on-site potential disorder where the nodal DoS change
is exponentially small in the inverse of the perturbation parameter.
Furthermore, as $n_{v}$ is increased, a wider symmetrical structure
emerges at base of the central peak, which signals that inter-vacancy
hybridization is turning the bound states into continuum resonances
and spreading their weight over a finite spectral region. Interestingly,
this energy spreading entails a finer structure of subsidiary peaks
that flank the node for dilute concentrations, $n_{v}\lesssim1\%$;
see Fig.\,\ref{fig-01}\,b. This comb of sharp resonances is characteristic
of three-dimensional\,(3D) WSMs and cannot be observed in the DoS
of a 2D Dirac semimetal with vacancies\,\citep{Pires2022}.

\noindent \emph{Magnetic response}.---We have argued that, in general,
quantum interference effects between vacancies lift the degeneracy
of zero-energy modes, thus generating sharp resonances shifted away
from the node. These hybrid states are no longer proper bound states
but, crucially, still retain a quasi-localized character. In Ref.\,\citep{Pires2022},
this picture was confirmed by exact diagonalization results, but here
we take a less direct but more practical approach. Rather than probing
the real-space wavefunctions, we scrutinize the modification to the
DoS induced by an external uniform magnetic field, in the presence
of vacancies. The rationale for this is that quasi-localized electronic
states should remain robust to applied magnetic fields, thus effectively
freezing out the DoS around the nodal energy. This intuition is backed
by our DoS 
\begin{figure}[t]
\begin{centering}
\hspace{-0.2cm}\includegraphics[scale=0.22]{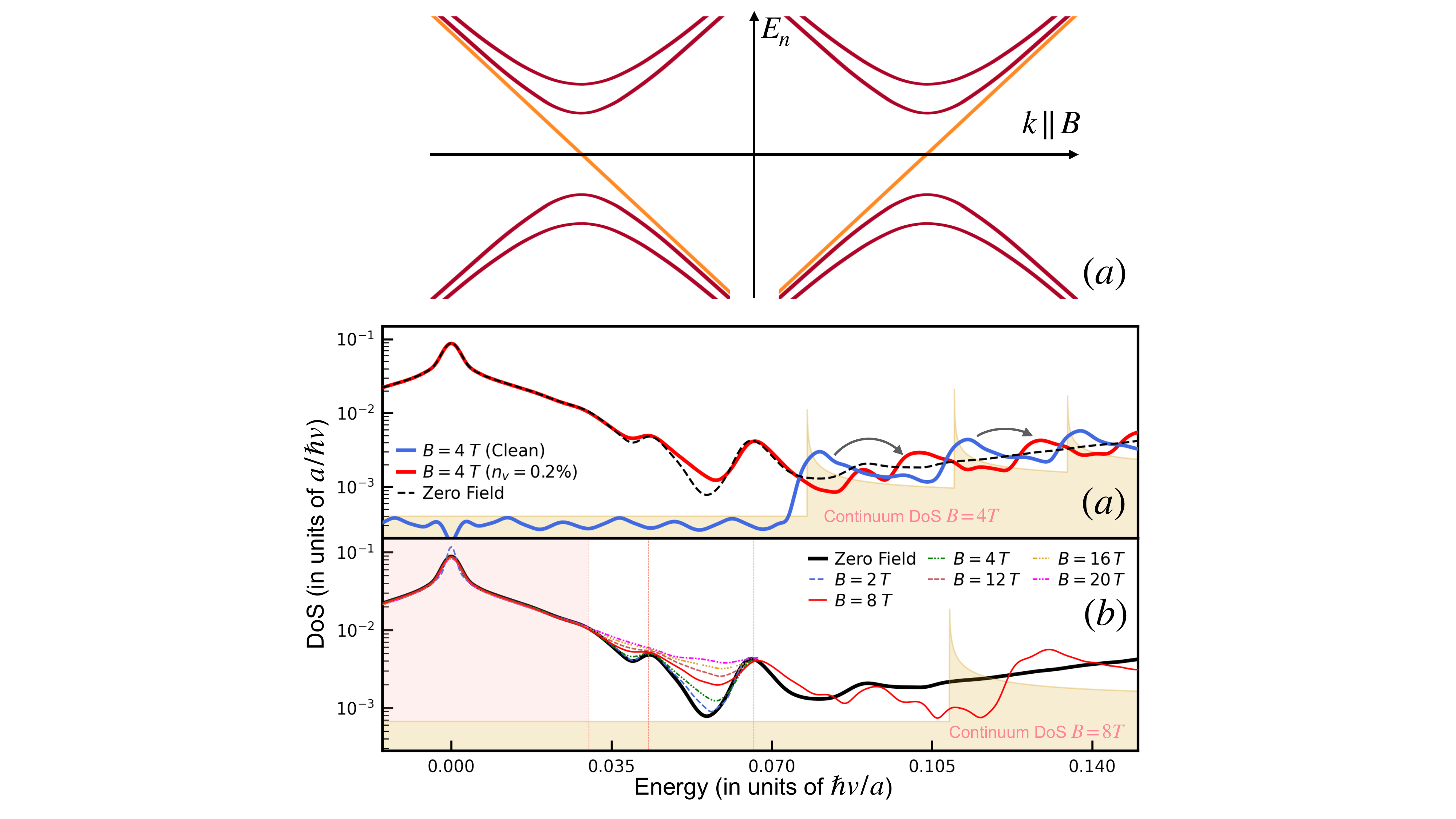}
\par\end{centering}
\vspace{-0.15cm}

\caption{\label{fig-02}\textit{(a) }DoS of a system with (red) and without
(blue) dilute vacancies subjected to a $4T$ magnetic field along
$\mathbf{z}$.\textit{ (b)} DoS with a vacancy concentration of $n_{v}\!=\!0.2\%$
for a selection of magnetic field strengths\textcolor{black}{. Insensitive
regions of the spectrum are marked or shaded in salmon. Above $E\!=\!0.07\,\hbar v/a$,
only the curves for $B=0T\text{ and }8T$ are represented. In both
cases, }the zero field curves are shown in black and the low-energy
theory DoS appear in the background, a lattice spacing of $a\!=\!5\,\mathring{\textrm{A}}$
was assumed, and the spectral resolution was set to $10^{-3}\,\hbar v/a$.}

\vspace{-0.3cm}
\end{figure}
 
\begin{figure}[h]
\begin{centering}
\hspace{-0.1cm}\includegraphics[scale=0.165]{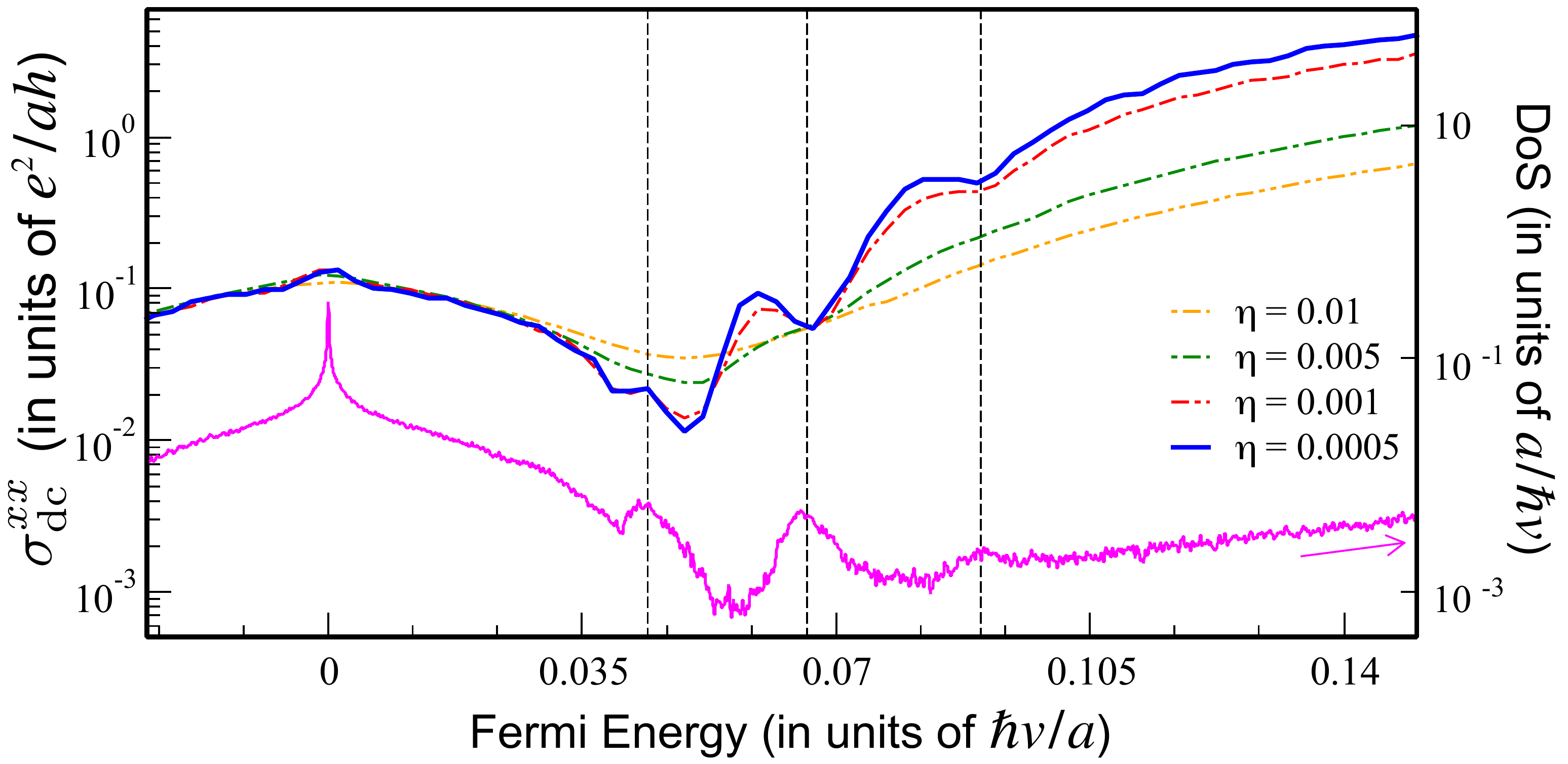}
\par\end{centering}
\vspace{-0.3cm}

\caption{\label{fig-03}Zero temperature bulk dc-conductivity as a function
of the Fermi energy for several values of the inelastic broadening,
$\eta$. The DoS is also depicted in magenta for reference, with the
vertical black lines indicating prominent subsidiary resonances. The
vacancy concentration was fixed to $n_{v}\!=\!0.2\%$.}

\vspace{-0.6cm}
\end{figure}

\noindent simulations for selected field strengths in the range $2\!-\!20$
T shown in Fig.\,\ref{fig-02}.\textcolor{black}{{} We also present
the DoS calculated in the clean system for a direct comparison at
a finite $B$. Two features are worth mentioning. First of all, near
the nodal energy, the DoS stays pinned to its zero-field values which
corroborates the quasi-localized character of defect states. Indeed,
the DoS central peak (shaded region in Fig.\,}\ref{fig-02}\,\textcolor{black}{b)
and the subsidiary resonances (vertical dashed lines) are nearly unchanged
by the external field. This contrasts with the regions in between
the subsidiary resonances, as shown in Fig.\,\ref{fig-02}\,b, and
also with the characteristic plateau-shaped DoS expected in a clean
Weyl node, whose $n\!=\!0$ (chiral) Landau level disperses linearly
along the magnetic field\,\citep{Nielsen1983,Shao16,Klier17}. Second,
the set of DoS peaks (van Hove singularities) at the bottom of the
non-chiral Landau levels is largely unaffected by the addition of
point defects, except for the global shift away from the nodal energy
as indicated by the arrows in }Fig.\,\ref{fig-02}\,\textcolor{black}{a.
This reinforces the notion that vacancies introduce new (quasi-localized)
states around zero energy, but trivially affect the electronic structure
far from it.}

\emph{Bulk transport and diffusivity}.---The DoS of a bulk 3D system
is not easily accessible to experiments, but it directly influences
a range of thermodynamic and transport properties. An experimentally
measurable quantity of great interest is the bulk dc-conductivity.
\textcolor{black}{In the diffusive regime, the latter is related to
the DoS at the Fermi level according to $\sigma_{\text{dc}}\!(E_{F})\!=\!e^{2}\rho(E_{F})\mathcal{D}(E_{F})$,
where $\mathcal{D}(E_{F})$ is the electronic diffusivity. Thereby,
the strong energy dependence imparted by vacancies on the DoS (}\textit{i.e}\textcolor{black}{,
the emergent nodal peak and the comb of subsidiary resonances; see
Fig.\,\ref{fig-01}) is expected to have a counterpart on the electrical
conductivity. To explore this further, we evaluate the $T\!=\!0$
linear dc-conductivity of systems with a finite vacancy concentration
using the Kubo-Greenwood formula\,\citep{Kubo57,Greenwood58}}

\vspace{-0.34cm}

\textcolor{black}{
\begin{equation}
\!\!\!\!\!\sigma_{\!\textrm{dc}}^{ii}\!\left(E_{{\scriptscriptstyle F}}\right)\!=\!\!\av{\!\frac{e^{2}h}{a^{3}L^{3}}\text{Tr}\!\left[\!\mathscr{V}^{i}\delta_{\eta}\!\left(E_{{\scriptscriptstyle F}}\!-\!\mathcal{H}_{v}\right)\!\mathscr{V}^{i}\delta_{\eta}\!\left(E_{{\scriptscriptstyle F}}\!-\!\mathcal{H}_{v}\right)\!\right]\!\!}\!,\!\!\!\label{eq:KuboGreenwood_Dirac}
\end{equation}
}

\noindent where $\mathcal{H}_{v}$ is the Hamiltonian with random
point defects, $\mathscr{V}^{i}=(i/\hbar)[\mathcal{H}_{v},\hat{x}_{i}]$
is the velocity operator along $\hat{x}_{i}$\,\footnote{We will drop the $i$ index, since the system has (on average) a cubic
symmetry.}, $\textrm{Tr}$ is the trace operation and $\av{\cdots}$ represents
the configurational average over the vacancy distribution. \textcolor{black}{The
numerical evaluation of Eq.\,(\ref{eq:KuboGreenwood_Dirac}) is carried
out with the efficient single-shot Chebyshev Polynomial Green's Function
method\,\citep{Ferreira2015,Joao2020}, and requires a broadening
of the Dirac-$\delta$ functions by a spectral resolution $\eta\gtrsim\delta\varepsilon$,
where $\delta\varepsilon$ is the mean-level spacing. We note that
the broadening parameter can alternatively be viewed as playing the
role of a}\textcolor{black}{\emph{ }}\textcolor{black}{phenomenological
self-energy with origin in electron-phonon scattering events\,\citep{Thouless1981}.
Because we are interested in disorder-limited charge carrier transport,
hereafter we focus on the behavior of $\sigma_{\textrm{dc}}$ in the
thermodynamic limit with $\eta\rightarrow\delta\varepsilon$. In Fig.\,\ref{fig-03},
we present the fully converged results of our Kubo simulations with
spectral resolutions down to $\eta\!=\!5\times10^{-4}\hbar v/a$.
Strikingly, these results show that the WSM dc-conductivity is strongly
non monotonic at low energies, which makes point defects strong contenders
for studies of quantum interference effects by means of standard electrical
transport measurements at zero magnetic fields. Indeed, the Fermi
level dependence of $\sigma_{\!\textrm{dc}}^{xx}$ is seen to nicely
track the oscillatory behavior of the DoS (see Fig.\,\ref{fig-03}),
confirming the intuition that the novel spectral fingerprints of point
defects are reflected on the linear-response properties. We expect
this oscillatory behavior in the dc-conductivity to remain visible
up to moderate temperatures (on the order of $100\,K$) as long as
the oscillation scale ($\sim0.01\,\hbar v/a$) remains sizable compared
to the thermal broadening (see SM\,\citep{SuppMat} for additional
discussions).}\textcolor{brown}{{} }\textcolor{black}{Additionally,
our quantum transport simulations reveal that the spectral convergence
rate (which measures how quickly the conductivity curves saturate
with decreasing $\eta$) is strongly dependent on the precise Fermi
level location in the low energy regime. We attribute this effect
to a strong energy dependence of the disorder self-energy and related
elastic scattering times\,\citep{Joao2021}. Indeed, the dips observed
at the resonant energies signal a }\textit{\textcolor{black}{strong
suppression of electron diffusivity. }}\textit{\textcolor{black}{\emph{This
is a signature of}}}\textcolor{black}{\emph{ }}\textcolor{black}{scattering
cross-sections that are enhanced at these energies\,\citep{Elattari2000}
by coherent multiple scattering processes among a few vacancies (see
also Ref.\,\citep{Pires2022}).}

\begin{figure}[t]
\vspace{-0.25cm}
\begin{centering}
\hspace{-0.2cm}\includegraphics[scale=0.23]{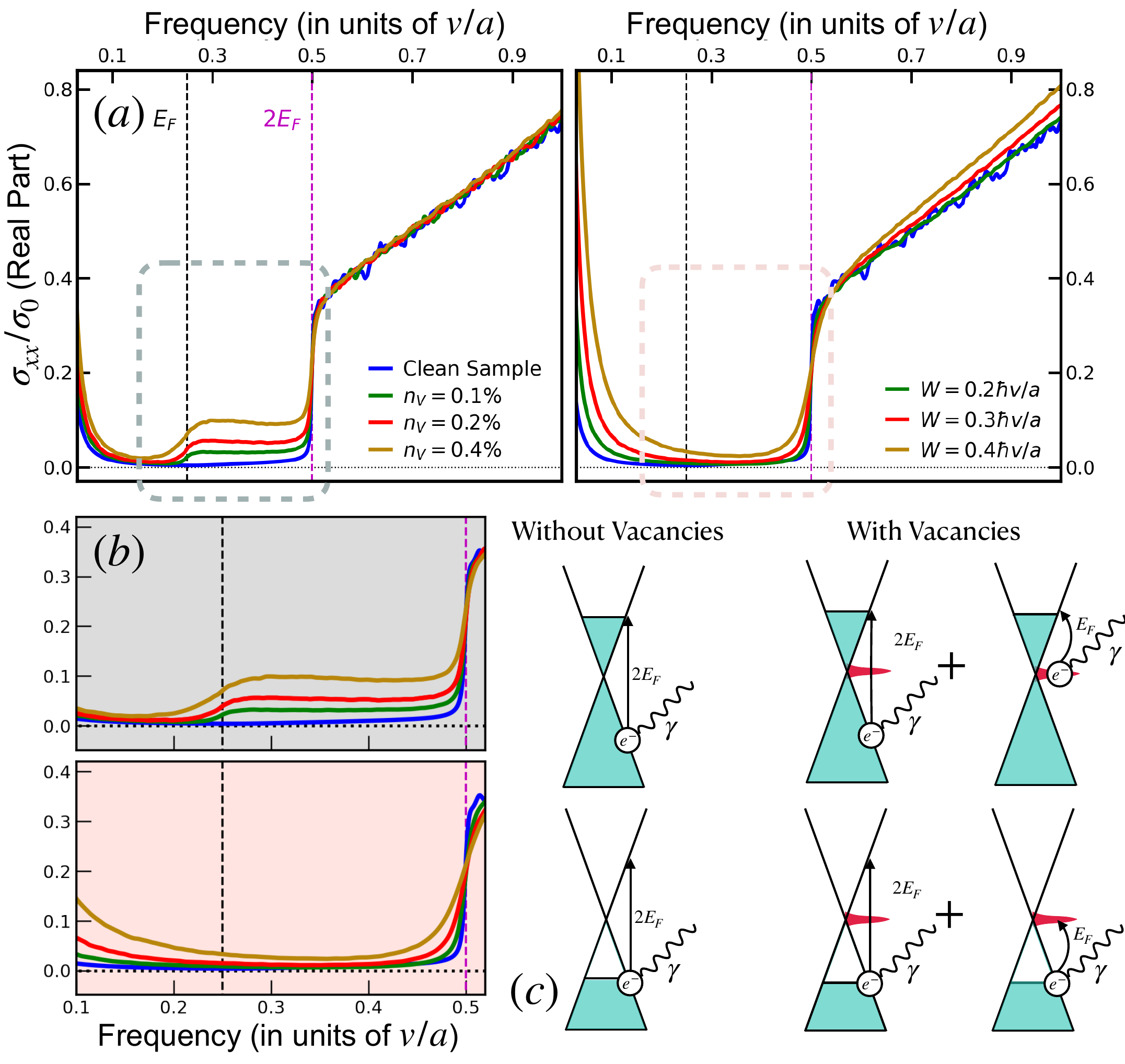}
\par\end{centering}
\vspace{-0.3cm}

\caption{\label{fig-04}\textit{\emph{(a)}}\emph{ }Longitudinal linear optical
conductivity of a doped WSM computed with an energy broadening $\eta\!=\!0.004\hbar v/a$
and a Fermi energy $\varepsilon_{F}=0.25\hbar v/a$ with a finite
vacancy concentration. The clean case is shown in blue. \textit{\emph{(b)}}\emph{
}Close-up of the vacancy-induced features in $\sigma_{xx}(\omega)$.\emph{
}\textit{\emph{(c)}} Relevant physical mechanisms for $\omega\!>\!\protect\abs{E_{\text{F}}}$.
Vertical arrows represent inter-band transitions between Bloch states,
while curved arrows indicate transitions between Bloch states and
nodal quasi-bound states. All plots are converged in the number of
polynomials.}

\vspace{-0.7cm}
\end{figure}

\emph{Optical response}.---Next, we search for the optical signatures
of t\textcolor{black}{he novel effects described above. Naively, the
existence of a macroscopic number of zero energy modes generated by
vacancy defects is expected to significantly change the allowed single-photon
optical transitions. For a pristine, but slightly doped WSM, vertical
inter-band transitions are Pauli blocked for $\omega\!<\!2\abs{E_{\text{F}}}$.
This defines the system's optical gap, where the optical conductivity
is mostly imaginary and reverses sign at $\omega\!=\!\abs{E_{\text{F}}}$.
In Fig.\,\ref{fig-04}\,a, the optical response of the clean WSM
is represented in blue, where a linear growth of the conductivity's
real part with $\omega$ can be seen above the optical gap\,\citep{Ashby14}.
This frequency dependence is a hallmark of 3D systems with linear
dispersion, differing from the universal plateau-shaped infrared conductivity
of graphene\,\citep{Nair08}.}

To account for the impact of randomly distributed vacancies, we have
calculated the linear optical conductivity by means of a kernel-convoluted
Chebyshev expansion of the finite-frequency Kubo formula\,\citep{Ferreira16,Joao19,Joao2020}.
Our findings are summarized in Fig.\,\ref{fig-04}. In stark contrast
to the standard situation without point defects, \emph{a finite dissipative
response now appears as a conductivity plateau in the interval}\textcolor{brown}{{}
}\textcolor{black}{$\abs{E_{\text{F}}}\!<\!\omega\!<\!2\abs{E_{\text{F}}}$.}\emph{
}We note that the plateau\emph{ }height grows quickly with the vacancy
concentration $n_{v}$, and thus should be observable under experimentally
realistic conditions. For completeness and contrast, we also show
the optical conductivity of the WMS lattice model with a box distribution
of on-site disorder (of strength $W$) and no point defects; see Figs.\,\ref{fig-04}\,a-b.
In this case, the uncorrelated random potential produces a conventional
broadening of the optical gap transition without specific features.

To explain the unusual optical response reported here, we note that
vacancies act as resonant defects that introduce a macroscopic number
of states near the band center of a time-reversal symmetric ($\mathcal{T}$-symmetric)
Weyl lattice, and that such states strongly break translation symmetry
owing to their quasi-localized character. Hence, the vacancy defects
efficiently participate in momentum non-conserving (but energy conserving)
transitions involving extended states at the Fermi energy; see schematic
in Fig.\,\ref{fig-04}\,c. These processes give a strong contribution
to the real part of the conductivity starting a\textcolor{black}{t
$\omega\!=\!\abs{E_{\text{F}}}$, }thus providing an unambiguous signature
of defect-induced nodal modes accessible to experiments. Interestingly,
the sharpness of transition to the vacancy-induced absorption regime
at $\omega=\varepsilon_{F}$ provides an estimate to the width of
the central peak in the DoS, and thus the extent of quantum-coherent
inter-vacancy hybridization.

\emph{C}onclusions and outlook.\,---Hitherto unknown effects of
vacancies in the electronic structure, charge carrier transport and
optical response of \textcolor{black}{a cubic }$\mathcal{T}$-symmetric
Weyl semimetal were investigated using real-space Green's function
calculations. The first new insight is the emergence of a strong enhancement
of the DoS in and around the nodal energy (see Ref.\,\citep{Pires2022}
for a complementary study of vacancy-induced bound states), which
is in contrast to the previously studied cases of non-resonant disorder.
As the defect concentration is increased, an efficient build-up of
inter-vacancy hybridization effects leads to a broadening of the nodal
DoS peak and, different from the established picture for 2D Dirac
semimetals, also gives rise to a symmetric comb of subsidiary resonances
at finite energies. The vacancy-induced states are quasi-localized
in real space, remain largely insensitive to applied magnetic fields
and possess low quantum diffusivity. \textcolor{black}{In addition,
these features are also robust to further perturbations\,\citep{Pires2022}
which hints that they will likely be present in real WSM crystals,
going beyond the model analyzed here.} 

Our main focus was on exploring experimentally
accessible routes to observe distinctive signatures of resonant scattering
by point defects in \textcolor{black}{cubic Weyl semimetals, a relevant
source of disorder even in high-quality single crystals such those
belonging to the cubic/tetragonal TaAs family}\textcolor{black}{\,}\textcolor{black}{\citep{Besara2016,Liu2016}.}\,Two
major transport effects amenable to experimental detection were predicted.
First, we showed that the Fermi-energy dependence of the low-temperature
dc-conductivity \textcolor{black}{is characterized by an unusual oscillatory
behavior }showing a series of dips whenever a subsidiary resonance
crosses the Fermi level. There, the effect of an enhanced DoS is compensated
by the strongly suppressed electron diffusivity due to the quasi-localized
nature of the states. We antecipate that such a strong Fermi level
dependence would be amplified in measurements of the thermo-electric
power, which is proportional to \textcolor{black}{$d\sigma(E_{F})/dE_{F}$
b}y Mott's formula\,\citep{Mott2012}. Finally, we showed that vacancy-induced
states endow the Weyl system with a unique optical response, \textit{i.e.},
the emergence of a plateau-shaped linear absorption below the inter-band
threshold. \textcolor{black}{Since it does not qualitatively depend
on the Fermi level, this} optical route is particularly appealing
for studying bulk samples where the carrier density is \textcolor{black}{notoriously}
difficult to control. Our findings spotlight the role of point defects
in amplifying quantum interference fingerprints in transport phenomena,
setting the stage for understanding the  interplay of topology and
resonant disorder in \textcolor{black}{generic} Weyl semimetals. 

\emph{Acknowledgements.}\,---\,J.\,P.\,S.\,P., S.\,M.\,J.,
B.\,A. and J.\,M.\,V.\,P.\,L. acknowledge support from the Portuguese
Foundation for Science and Technology (FCT) within the Strategic Funding
UIDB/04650/2020, and through projects No.\,POCI-01-0145-FEDER-028887
(J.\,P.\,S.\,P., S.\,M.\,J. and J.\,M.\,V.\,P.\,L.) and No.\,CEECIND/02936/2017\,(B.\,A.).
J.\,P.\,S.\,P. and S.\,M.\,J are funded by FCT grants No.\,PD/BD/142774/2018
and PD/BD/142798/2018, respectively. A.\,F. acknowledges support
from the Royal Society (London) through a Royal Society University
Research Fellowship. The large-scale calculations were undertaken
on the HPC Viking Cluster of the University of York. We thank J. Dieplinger,
J.\,M.\,B Lopes dos Santos and A. Altland for fruitful discussions,
and David T.\,S. Perkins for proof-reading the manuscr\textcolor{black}{ipt.
We also thank the anonymous referees for providing valuable feedback.}

\vspace{-0.55cm}

\bibliographystyle{apsrev4-2}
\bibliography{References}

\clearpage
\onecolumngrid

\section*{Supplementary Material}

\subsection*{Additional Numerical Results}

\begin{figure}[h]
\includegraphics[scale=0.38]{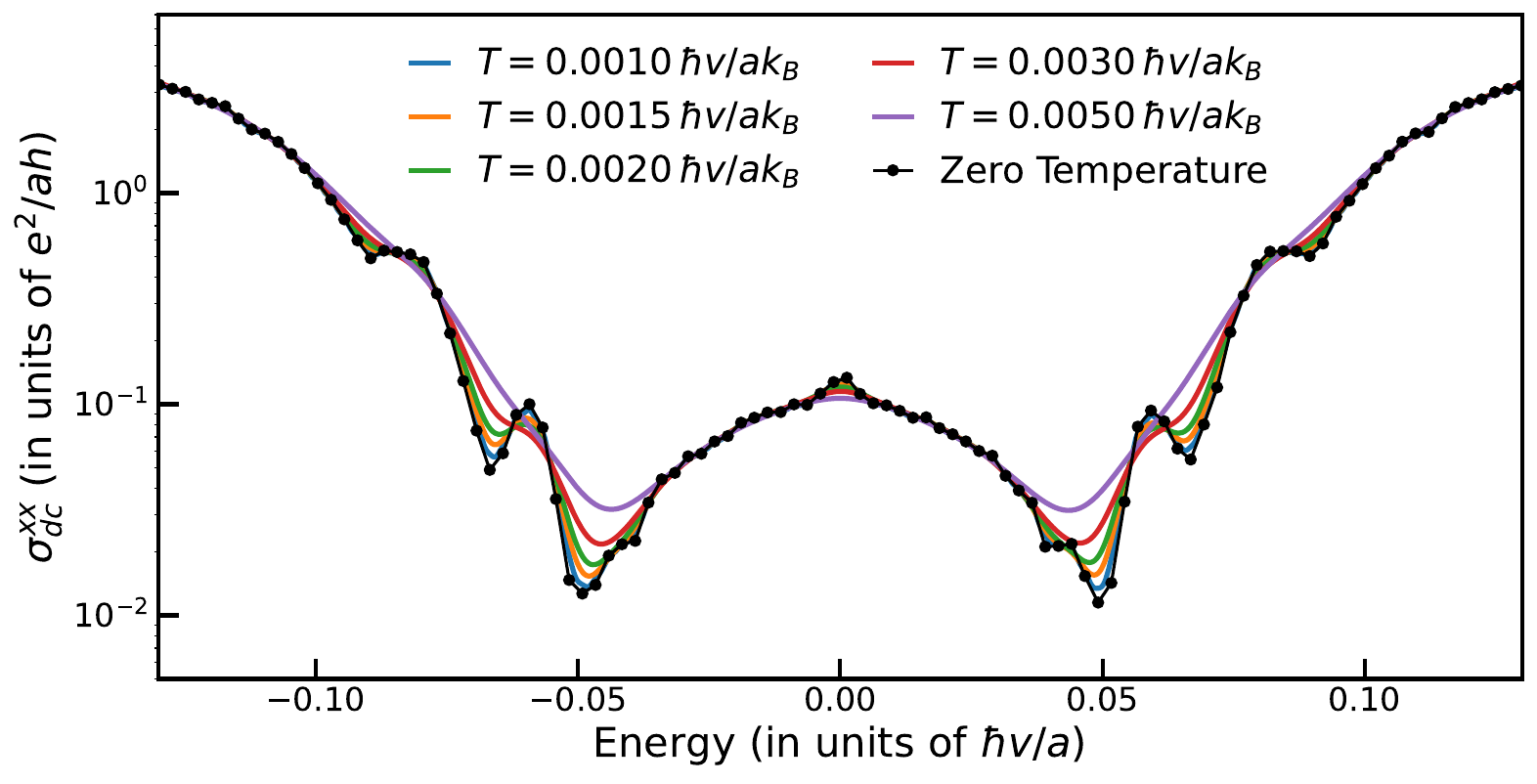}

\vspace{-0.3cm}

\caption{\label{fig:FiniteT_Cond}Bulk dc-conductivity as a function of the
Fermi energy for several values of temperature ($T$), and using the
smallest inelastic broadening shown in Fig.\,3 of the main text,
$\eta=0.0005\hbar v/a$. For reference, we note that $T=0.001\hbar v/ak_{B}\sim100K$.}

\vspace{-0.3cm}
\end{figure}

For completeness, we include here a further results on the dc-conductivity
of the system calculated at finite temperatures, as well as the system's
full optical response. In Fig\,\ref{fig:FiniteT_Cond}, we show the
same results shown in Fig.\,3 of the main text but now calculated
at finite temperatures. These plots were obtained by convoluting the
zero temperature data with a kernel given by the energy-derivative
of the Fermi-Dirac distribution, i.e. $-df_{\text{F}}(E,T)/dE=\left[2k_{B}T+2k_{B}T\cosh\left(E/k_{B}T\right)\right]^{{\scriptscriptstyle -1}}$. 

\begin{figure}[h]
\includegraphics[scale=0.26]{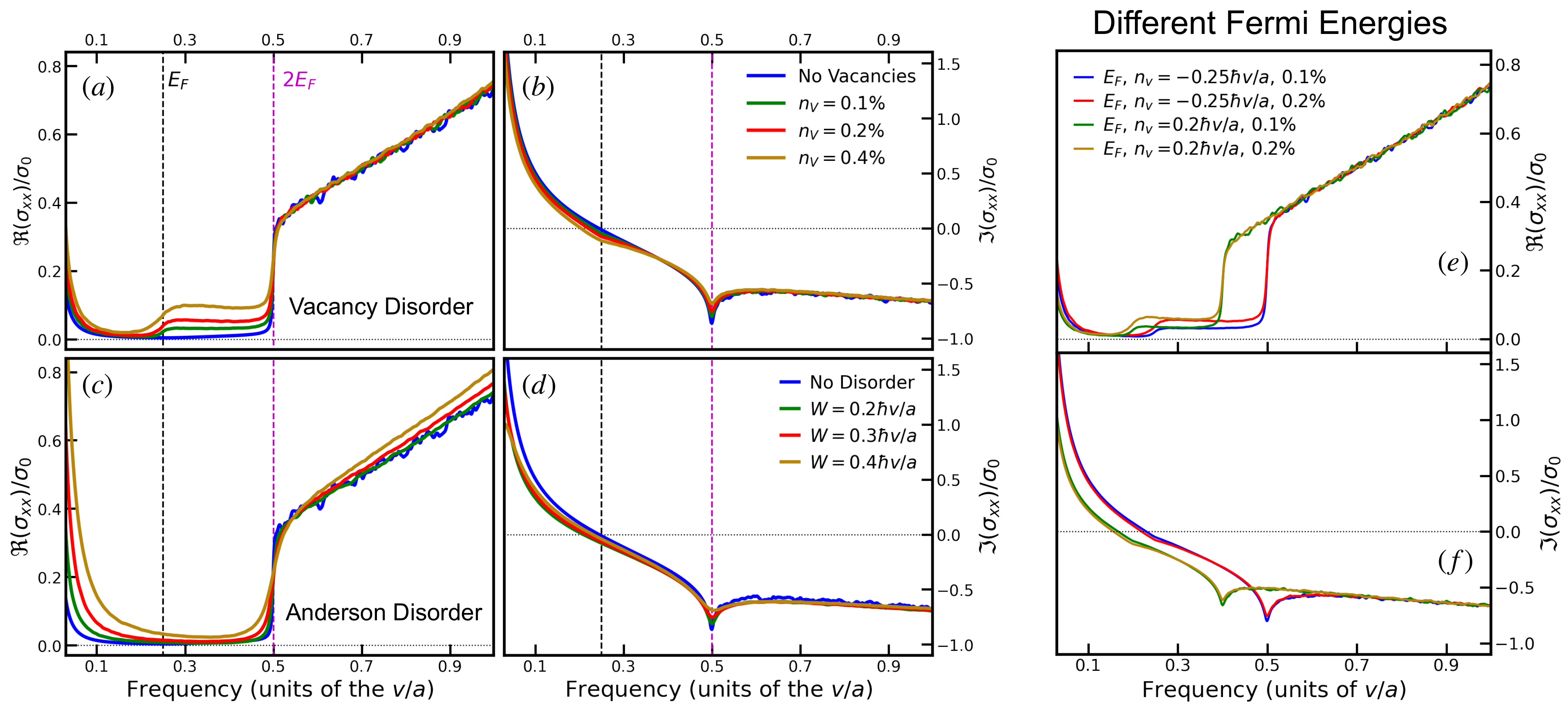}

\vspace{-0.3cm}

\caption{\label{fig:OpticalResponse}Additional numerical results for the real
and imaginary parts of the linear optical conductivity of the $\mathcal{T}$-symmetric
Weyl semimetal in the presence of random point-like vacancies {[}(a),
(b), (e) and (f){]} with a concentration $n_{v}$ and on-site Anderson
disorder {[}(c) and (d){]} of strength $W$. The plots of (a)-(d)
were obtained with a Fermi level placed at $E_{F}\!=\!0.25\hbar v/a$,
while the results of (e) and (f) considered alternative fillings of
$E_{F}=-0.25\hbar v/a$ and $E_{F}=0.2\hbar v/a$.}

\vspace{-0.3cm}
\end{figure}

In Fig.\,\ref{fig:OpticalResponse}\,(a)-(d), we present the real
and imaginary parts of the linear optical conductivity in the presence
of vacancies {[}(a) and (b){]} and on-site Anderson disorder {[}(c)
and (d){]}. Besides the effects on the real component already discussed
in the main text, one also observes that the imaginary parts are only
slightly changed relative to the clean system's response. The only
notable features is slight a downshift of the frequency at which it
reverses sign, something that happens in both models of disorder.
In Fig.\,\ref{fig:OpticalResponse}\,(e)-(f) the same representation
for vacancies is made, considering different values of the Fermi energy.
The results are qualitatively similar.

\end{document}